\documentclass[journal]{IEEEtran}

\usepackage{amsmath}
\usepackage{graphicx}
\usepackage{amssymb}
\usepackage{mathrsfs}
\usepackage{mathtools}
\usepackage{makecell,multirow}

\usepackage[ruled,vlined]{algorithm2e}

\usepackage{bm}
\usepackage{stfloats}

\usepackage{cite}

\usepackage{subfigure}
\usepackage{color}

\newcommand{\MyAlterDel}[1]{}

\usepackage{booktabs}
\usepackage{multirow}
\usepackage{array}

\usepackage{flushend}

\usepackage{url}

\begin{document}
\bstctlcite{Ref-MEC-TWC:BSTcontrol}
\makeatletter
\def\bstctlcite{\@ifnextchar[{\@bstctlcite}{\@bstctl
cite[@auxout]}}
\def\bstctlcite[#1]#2{\@bsphack
\@for\@citeb:=#2\do{%
\edef\@citeb{\expandafter\@firstofone\@citeb}%
\if@filesw\immediate\write\csname #1\endcsname{\s
tring\citation{\@citeb}}\fi}%
\@esphack}
\makeatother

\title{Resilience in Industrial Internet of Things Systems: A Communication Perspective}

\author{Hao Wu, Yifan Miao, Peng Zhang, Yang Tian, Hui Tian
%\vspace{-2em}
\iffalse
Hui Tian, \IEEEmembership{Senior Member, IEEE}, Shaoshuai Fan, and Jiazhi Ren
\vspace{-2em}% <-this % stops a space
\thanks{This work was supported in part by the State Grid Corporation of China Major Project, Research on Key Technologies of Electricity Distribution Internet of Things, under Grant 5400-201955452A-0-0-00, and in part by the Beijing University of Posts and Telecommunications (BUPT) Excellent Ph.D. Students Foundation under Grant CX2019108. Paper no. TII-19-5268. \emph{(Corresponding author: Hui Tian)}
}
\fi

\thanks{H. Wu is with the State Key Laboratory of Networking and Switching Technology,
Beijing University of Posts and Telecommunications, Beijing 100876, China, and also with the Center for Complex Network
Research, Northeastern University, Boston, MA 02115, USA (e-mail:
wh9405@bupt.edu.cn)}

\thanks{Y. Miao, P. Zhang, Y. Tian, and H. Tian are with the State Key Laboratory of Networking and Switching Technology,
Beijing University of Posts and Telecommunications, Beijing 100876, China (e-mail: tianhui@bupt.edu.cn; zpeng3@bupt.edu.cn; tianyang@bupt.edu.cn; miaoyifan@bupt.edu.cn).}

}

\maketitle

\begin{abstract}
Industrial Internet of Things is an ultra-large-scale system that is much more sophisticated and fragile than conventional industrial platforms.
The effective management of such a system relies heavily on the resilience of the network, especially the communication part.
Imperative as resilient communication is, there is not enough attention from literature and a standardized framework is still missing.
In awareness of these, this paper intends to provide a systematic overview of resilience in IIoT with a communication perspective, aiming to answer the questions of \textit{why} we need it, \textit{what} it is, \textit{how} to enhance it, and \textit{where} it can be applied.
Specifically, we emphasize the urgency of resilience studies via examining existing literature and analyzing malfunction data from a real satellite communication system.
Resilience-related concepts and metrics, together with standardization efforts are then summarized and discussed, presenting a basic framework for analyzing the resilience of the system before, during, and after disruptive events.
On the basis of the framework, key resilience concerns associated with the design, deployment, and operation of IIoT are briefly described to shed light on the methods for resilience enhancement.
Promising resilient applications in different IIoT sectors are also introduced to highlight the opportunities and challenges in practical implementations.
\end{abstract}


\vspace*{-10pt}
\section{Introduction}
\IEEEPARstart{I}{ndustrial} Internet of Things (IIoT) has been regarded as a revolutionary approach to drive industrial efficiency, productivity, and performance to an unprecedented level~\cite{bibli:IIoT3,bibli:IIoT4,bibli:myIoT}.
Unlike conventional industrial platforms which are usually small-scale and independent of each other, IIoT is anticipated to be an ultra-large-scale system, notably based on advances in communication technologies~\cite{bibli:myTII}.
Such a system will be hard to coordinate efficiently and control effectively without in-built resilience.
Consequently, the resilience of communication design is of significant importance and could spawn an entirely new branch of research to achieve the full potential of IIoT.
%It is anticipated that, by the year 2024, the IIoT market size will reach to \$197 billion.
%It promises ubiquitous interaction between the physical world and its digital counterpart, bringing the to
%As a system of systems, IIoT consists of different types of sub-networks such as power grid, communication system, and control network, among which the communication system arouses most concerns towards engineering of the future industrial environments.
%To achieve the full potential of IIoT communication, the resilience of engineering design is of prime importance.
%In the past, communication technologies in industrial environments were mostly based on
%Although interweaved with other networks, communication system, working as the commander to coordinate with diverse components of IIoT, is actually the first step towards engineering of the future industrial systems.
%The full potential of IIoT communication has yet to be reached considering challenges posed by those different research fields, among which communication arouses most concerns.
%Given its commander role in coordinating diverse components of IIoT, communication is actually the most important part in intelligentizing
%first step towards engineering of the future industrial systems, highly promising

Although resilience has been carefully studied in various engineering domains, there is still no widely accepted definition of it.
Generally, it is believed that a resilient system should be able to resist undesirable events arising from natural or human interventions, adapt its function to constant stress or uncertainties, and recover quickly to a certain level of availability after disasters~\cite{bibli:SDN,bibli:roadmap,bibli:biblio}.
According to the definition, resilience is an overarching concept that covers the interests of different fields such as resistance, fault-tolerance, redundancy, robustness, recoverability, reliability, and so on.
In this paper, we adopt this generic definition of resilience as it is also consistent with the concept defined by the European Network and Information Security Agency (ENISA)~\cite{bibli:ENISA}.
Related concepts such as fault-tolerance will be used interchangeably with resilience when referring to specific scenarios in the sequel.
With respect to IIoT communication, system resilience against all kinds of unfavorable situations can be viewed from three aspects: extraction, manufacturing, and service~\cite{bibli:IIoT4}.
Regarding the extraction aspect, it is necessary to improve the system tolerance against faults that happened at the collection, transmission, and process progresses by adding a certain redundancy in information.
Then this information can be used to guide the manufacturing of IIoT.
The infrastructure used for manufacturing is usually vulnerable to unexpected disturbances and prone to unwanted faults.
Therefore, communication in IIoT needs to enable fast fault detection and quick restoration technologies to ensure the continuity of services.
All these proactive or reactive methods, however, rely on the design of a long-term resilient framework.

The design, deployment, and operation of resilience in IIoT pose great challenges such as standardization, scalability, interoperability, and interdependence.
As discussed, a lack of well-defined definitions and corresponding measuring metrics is one of the major challenges facing the standardization of IIoT resilience.
Additionally, the resilience requirements of various systems could be totally different.
There would not be a one-size-fits-all model that could be applicable for those distinct environments.
Each resilient system needs to be meticulously tailored depending on its own specific conditions.
As the development of IIoT gathers pace, heterogeneous infrastructures and sensors of different generations in the system would further exacerbate the difficulty for interoperability.
Meanwhile, due to the complicated interconnectivity and interdependence between functionally different components, IIoT systems would become increasingly integrated, making the traditional isolated resilience analysis inapplicable.

Resilience is actually a popular topic in the area of complex science systems~\cite{bibli:CES,bibli:biblio}, such as computer science, ecology, and environmental science.
However, these works could not be easily extended to the communication scenario in IIoT, where infrastructure is highly coupled temporally and spatially.
While there are several reviews of definitions or use cases over communication in the Internet of things (IoT) or cyber-physical system (CPS), the majority of them focus on one building block of a resilient system~\cite{bibli:SDN,bibli:roadmap,bibli:extended,bibli:attack,bibli:cascade}.
For instance, the authors of~\cite{bibli:extended} paid attention to the security problem in the scenario of mobile edge computing and fog.
In~\cite{bibli:attack}, Daniel \textit{et al} considered the cyber-attack condition, which is only one type of abnormal event happening in communication systems.
Similarly, studies in~\cite{bibli:cascade} provided a comprehensive review work over a type of large system damage, cascading failures.
Efficient cascading failure models and reliability analysis methods, together with mitigation strategies are presented to facilitate quantitative analysis.
All of the above works provide interesting aspects for the research of IIoT resilience, however, a systematic study emerging from standardization efforts is still missing.
%Most recently, Liudong et al paid their attentions to another type of large system damage - cascading failures.

Motivated by the limitations of existing literature, this paper addresses resilience in IIoT from a communication perspective, intending to provide a systematic overview of and arouse more interest in the area.
Specifically, we aim to answer the questions of \textit{why} we need resilience, \textit{what} it is, \textit{how} to enhance it, and \textit{where} it can be used.
We highlight the necessity and significance of resilient communication through the analysis of bibliography and real malfunction data of satellite base stations.
As a widely accepted definition of resilience is unavailable, we clarify the different concepts related to resilience and elaborate corresponding metrics for quantitative analysis.
The concerns for building resilience associated with the design, deployment, and operation of IIoT systems are then discussed, accompanied by some solving strategies.
With these strategies, we further shed light on the benefits brought by resilient communication in three promising IIoT sectors.
%To the best of our knowledge, it is the first attempt to discuss IIoT resilience with both views from ongoing academic studies, industrial implementations, and standardization activities.

Distinct from existing works, this article represents the first attempt to discuss IIoT resilience with both views from ongoing academic studies, industrial implementations, and standardization activities.
It provides a systematic overview of state-of-the-art efforts and reveals the huge potential of resilience research.
Works in the areas of concept taxonomies, quantifiable metrics, enhancement technologies, and possible research directions are put together under a single umbrella of resilient communication.
It intends to highlight the imperativeness of a communication perspective for resilience study, provide a source of information for researchers interested in the field, and stimulate more in-depth work on the subject.

The rest of this paper is organized as follows.
Section II explains the imperativeness of resilience research from a communication perspective.
Then, section III provides an overview of related resilience concepts and metrics, as well as standardization efforts.
Next, key concerns for achieving resilience in IIoT systems are described in Section IV.
Some promising resilient applications from the communication perspective are then presented in Section V.
Finally, we give some concluding remarks in Section VI.

\section{Imperativeness of a Communication Perspective for Resilience}
In this section, we give a holistic view of current literature on engineering resilience via bibliometric analysis.
It shows that studies on resilience are often restricted to specific research areas, with the communication perspective often overlooked.
Could it be that resilience in communication systems is not that important?
The answer is definitely no.
A nine-month record of the operating status of satellite base stations in two provinces of China reveals that real communication systems are actually prone to failure and the characteristics of malfunctions are more complicated than we used to believe.
The details are illustrated in the following.
%A communication perspective of resilience research is like a gold mine that waiting for us to mine.

\subsection{Bibliometric Analysis}
\begin{figure}
\centering{\includegraphics[width=0.48\textwidth]{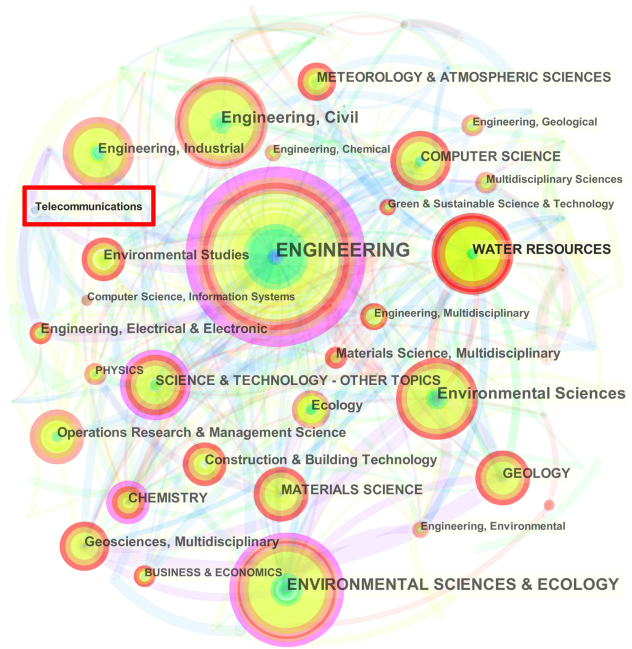}}
\caption{Cocitation map of 2388 articles with keywords searching of "resilience" and "engineering" in Web of Science database, created by CiteSpace.}
%\vspace*{-15pt}
\label{fig:ref}
\end{figure}

With keywords \textit{resilience} and \textit{engineering}, our literature search of \textit{Web of Science} database\footnote{The database (\url{https://webofknowledge.com/}) is chosen since it provides easy-to-access application programming interfaces and the most in-depth citation by source compared with other databases like Google Scholar and Scopus.}, identifies 2388 qualified articles from 2000 to March 2021.
The subject areas of these articles, as classified by the database, are then visualized and analyzed through CiteSpace~\cite{bibli:biblio}.
As illustrated in Fig.~\ref{fig:ref}, the size of a cluster reflects the number of papers found in that discipline.
The link between two different clusters represents the number of cocitations of papers in the two subjects.
Meanwhile, the brighter a cluster or link is, the more recently published papers they own.
Surprisingly, on the map, there are only 65 articles related to telecommunication as shown in the red frame.
What's worse, most of them were published before 2010, suggesting the absence of enough interest from the communication community in recent years.
However, like most engineering infrastructure systems, the communication network is very likely to suffer from continuous failures, which indeed requires more effort for resilience analysis.

\subsection{Urgency for Resilience Analysis}

\begin{figure}
\centering
\subfigure[]{
\label{fig:number} %% label for first subfigure
\includegraphics[width=0.4\textwidth]{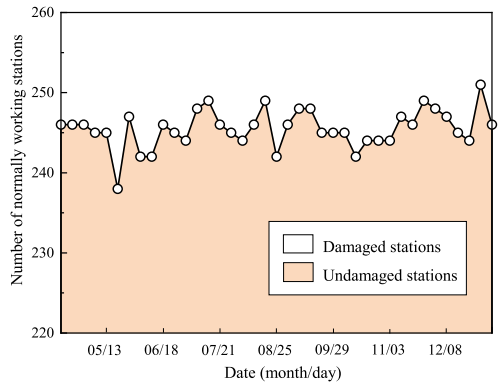}}
%\hspace{-0.05in}
\subfigure[]{
\label{fig:distribution} %% label for second subfigure
\includegraphics[width=0.48\textwidth]{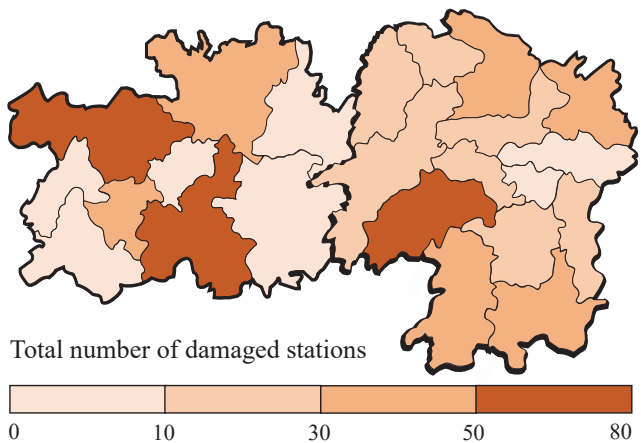}}
\caption{Recorded damaged satellite base station number within nine months in Guizhou and Hunan provinces of China. (a) The varied number of normally functioning base stations during observation. (b) The geographical distribution of damaged base stations, divided by cities.}
\label{fig:data} %% label for entire figure
\end{figure}

To evaluate the resilience of communication infrastructure, we tracked the data every seven days from a live status map of satellite base stations in Hunan and Guizhou provinces, China\footnote{\url{https://www.sobds.com/index.php?s=/Home/cors/station.html}}.
From 2020 April to December, more than 500 faulty cases were reported, together with their precise location, start time, and end time.
Over the observed time period, there is not even one time when all the 260 base stations are working normally, as shown in Fig.~\ref{fig:number}.
In other words, failures of infrastructure in a communication network are pretty prevalent in daily life.
Based on the locations of all the recorded failure cases, Fig.~\ref{fig:distribution} illustrates the number of failures that occurred in each city.
In contrast to the conventional random failure model~\cite{bibli:random}, the observed malfunction of base stations shows apparent geographical characteristics, implying that our understanding of communication systems' resilience is far from satisfactory.

To sum up, little resilience-related attention has been paid to communication systems, which, however, are more vulnerable and complicated than we used to believe.
The communication perspective of resilience is like a gold mine that is waiting for us to exploit, especially in the era of IIoT.

\section{Resilience in IIoT}
Although there are several attempts to define ``resilience" in communication, there is still no widely accepted definition.
Generally, it is believed that resilience is a function of time and runs through all the stages before, during, and after a disruptive event.
A comprehensive description of all related concepts in such a long time span is pretty tough.
However, some foundational aspects are worth highlighting.
%abnormal but legitimate traffic load, accidents and human mistakes, large-scale disasters, malicious attacks, occasional failures

\subsection{Related Resilience Concepts}
\begin{figure}
\centering{\includegraphics[width=1\linewidth]{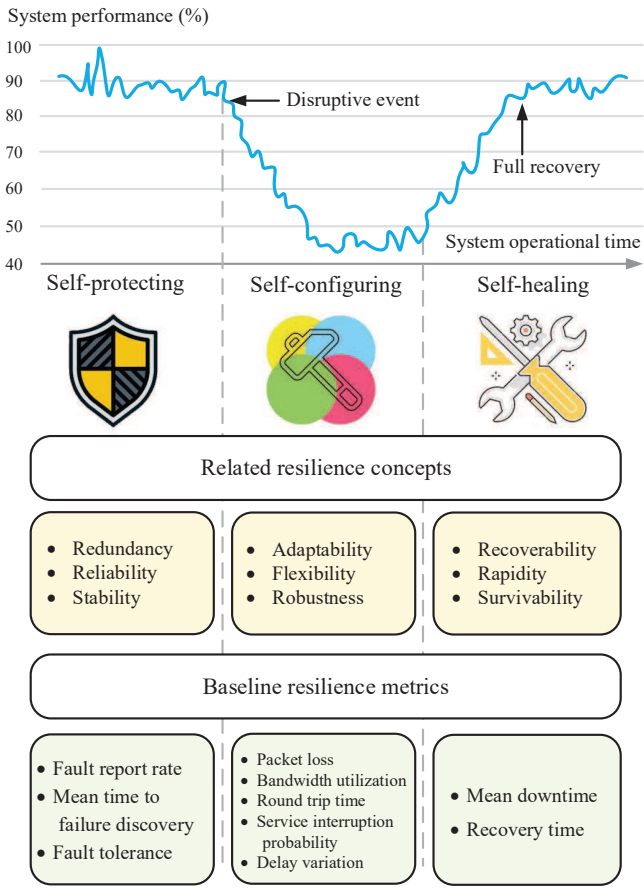}}
\caption{Operating states of the system when hit by a disruptive event, as well as resilience-related concepts and baseline metrics.}
%\vspace*{-15pt}
\label{fig:classification}
\end{figure}

Resilience is a superset of a large number of closely related concepts that have special angles of interest.
Depending on the appearance phase and impacts of these terms, we can roughly classify them into notions of self-protecting, self-configuring, and self-healing.
Referring to Fig.~\ref{fig:classification}, normally, an IIoT system works at a high-performance level and is able to resist minor external perturbations and internal failures, thanks to its embedded self-protecting mechanisms.
Once a disruption event that exceeds the system's protection capacity hits the network, system performance begins to fall, during which self-configuring is activated to maintain the function of the rest of the undamaged infrastructure.
After the detrimental event gets controlled, self-healing approaches are adopted to help the system recover back to its normal state.
Roughly speaking, all the other related resilience concepts can fall into at least one of the three notions.
Some may even fit two or all of them.
However, to simplify the classification and make it clearer, only the most relevant concepts are introduced in each category, as shown in Fig.~\ref{fig:classification}.
%Resilience incorporates a diversity of overlapping methods, policies, strategies or services to accomplish objects and fulfill purposes.

\subsubsection{Self-protecting (before collapse)}
IIoT systems are designed for self-protecting by using prognosis and redundancy.
Through excess capacity and backup components, IIoT is able to absorb disturbances and withstand potential damage, making them less likely to spread far and wide.
Reliability and stability are two concepts linked to maintaining core functionality performance in the event of disturbances.
They incorporate the concept of redundancy and put emphasis on the ability of systems to detect a threat earlier and quantify the severity of its impact.
Assisted by the advances in deep learning, prior system knowledge can be trained to evaluate the system vulnerability and considerably prevent the system from collapsing.

\subsubsection{Self-configuring (during collapse)}
IIoT employs self-protecting mechanisms only when there is an action in its predefined operational strategy set.
However, IIoT is evolving and pretty dynamic.
Most of the detrimental events cannot be anticipated and addressed during the design stage.
The system would reach its regulation limitation in the wake of excessive stress and be dangerous if suitable actions are not taken promptly, albeit the extreme events are of low probability.
Given all the above issues, IIoT needs to be capable of adaptability, which involves systems' inherent flexibility and enables spontaneous decision making and resource scheduling.
In this way, it can maintain the robustness of the existing functional infrastructure and protect the system from complete collapse.

\subsubsection{Self-healing (after collapse)}
Recoverability describes how quickly a system is able to restore and bounce back to its normal state.
The rapidity of recoverability heavily depends on the ability of the system to correctly receive information and respond to crises.
Nevertheless, under some circumstances, the system performance is only able to return to an acceptable level that is much less than its pre-disturbance condition.
Those circumstances measure the survivability of IIoT in the face of limited repairing resources and irreparable assets, which is also worthy of meticulous consideration during the design process.

%https://homo-digitalis.net/the-resilient-factory/

\subsection{Baseline Resilience Metrics}
Metrics are effective tools for engineers and managers to assess and regulate the resilience of IIoT systems.
However, there is still no widely accepted standardized metric that is used by stakeholders, since resilience itself is not a well-defined term.
This subsection will describe a number of quantifiable metrics for resilience measurements and provide a taxonomy overview.
In accordance with the temporary related classification in related concepts, baseline resilience metrics for communication can fall into three categories: self-protecting phase, self-configuring phase, and self-healing phase.

As shown in Fig.~\ref{fig:classification}, fault report rate and mean time to failure discovery in the self-protecting phase describe how many failures in the network can be detected and how fast the detection could be reported, respectively.
Fault tolerance is the limit of variation before the service level changes from the normal state to degraded.
They are all used for preparing the system to cope with faults and challenges.
During the self-configuring phase, the network relies on efficient system scheduling to make the best use of available resources to meet the communication demand as much as possible.
Performance-based metrics such as packet loss, bandwidth utilization, round trip time, service interruption probability, and delay variation are appropriate for this phase in measuring the impact of the disruptive event on the system's operational availability.
Although some of those metrics in the former two phases can be used for system recovery, the self-healing phase pays more attention to the mean downtime of infrastructure (equal to recovery time needed) from a more macroscopic view.

It is clear that an individual metric is not enough when demanded resilience measuring has different levels of abstraction.
One typical example is when a factory wants to understand the resilience status of the entire entity on a sector-wide basis.
More complicated and combined metrics will be required to achieve this, among which economic indicators are the ideal choices.
However, the complexity for analysis would increase rapidly as more baseline resilience metrics are aggregated.

\subsection{Standardization of Resilience}
%标准化的好处
Establishing a universally applicable standard for resilience facilitates the basic design, configuration, and operation of IIoT systems.
Resilience standardization efforts in communications have already existed and are being updated to accommodate the growing demand for evolving systems.
The first systematic one can be traced back to the technical report by the ENISA~\cite{bibli:ENISA}.
Focusing on public communications networks, ENISA provides an overview of related research projects and highlights the challenges in designing a resilient measurement framework.
%More specifically, a white paper concerning achieving end-to-end resilience through planned combination of prevention, protection, response and recovery is published to cope with

For the high-level discussion of communication resilience technologies, the most significant and leading efforts are the work carried out by the ITU-T~\cite{bibli:ITU}.
It identifies the systems and components that need common specifications, and describes the requirements for resilience enhancements, laying the foundation towards the next steps in standardization.
Besides, it also emphasizes the significance of disaster relief and gives resilience insights from a recovery perspective.
%It is worth noting that, the ITU-T technical report is the only standard originated from disaster relief and world-wide accepted standard.

Standardization activities involving resilience in IIoT first gained traction from Object Management Group (OMG) in 2015 and then listed as one of the key system concerns by the Industrial Internet Consortium (IIC)~\cite{bibli:IICk}.
For implementation considerations, IIC pays more attention to the security of networks, stating that communications between units must be validated before getting trusted.
As the standardization of IIoT gathers pace, the one Machine-to-Machine Partnership Project (oneM2M) then joined IIC's efforts in advancing the IIoT and reemphasized the importance of resilience in IIoT systems~\cite{bibli:IICa}.
The standardization process is still ongoing and requires more participation from related organizations and companies.

\section{Key Resilience Concerns}
IIoT resilience is technically complicated to achieve as IIoT is an interdependent network composed of a large number of elements depending on one another and is prone to cascading failures.
There is a plethora of resilience concerns connected with the design, deployment, and operation of IIoT systems.
We briefly describe some major ones in this section.

\subsection{Architecture Design}
\begin{figure}
\centering{\includegraphics[height=0.6\textwidth]{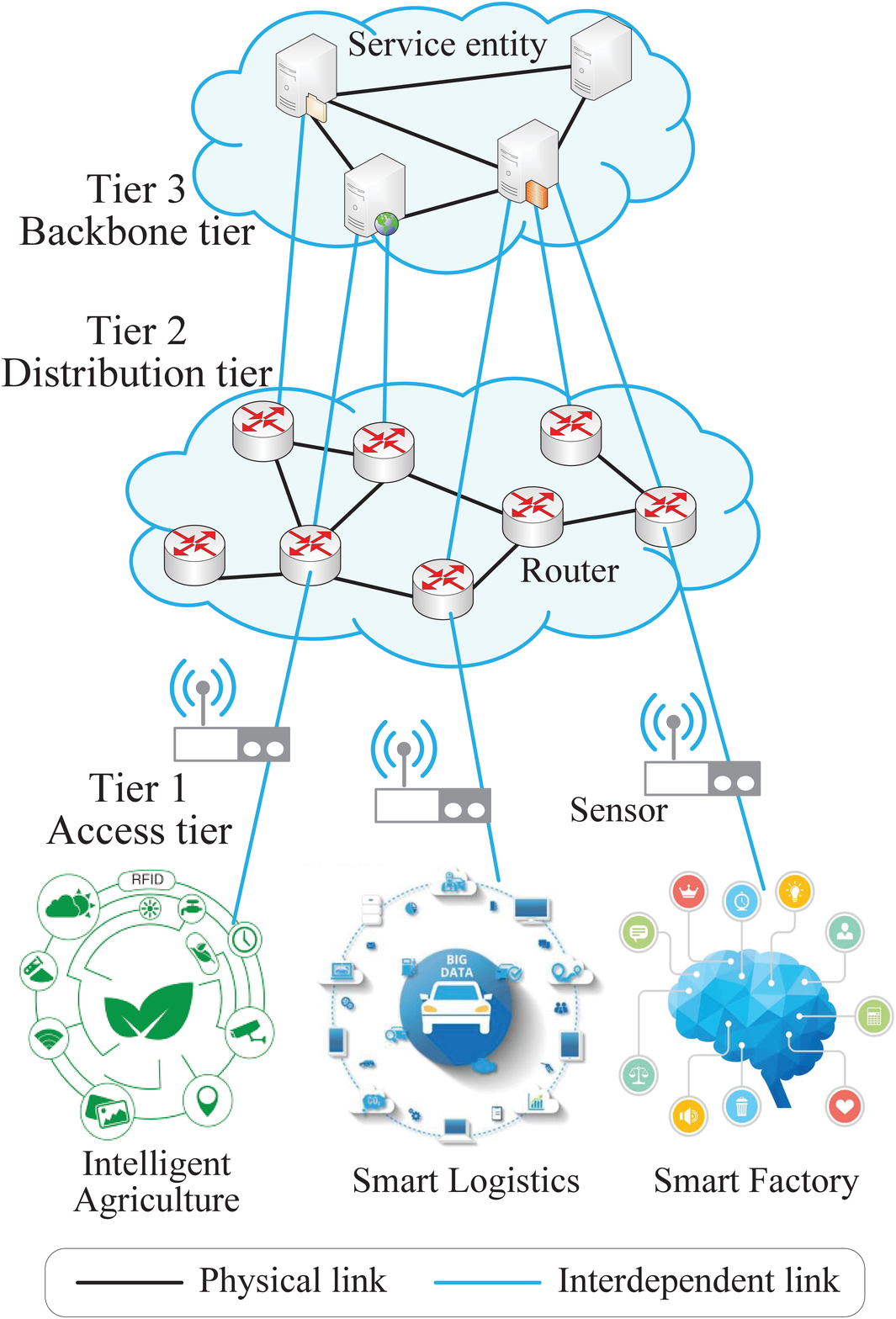}}
\caption{Hierarchical architecture of IIoT systems with backbone, distribution, and access tiers.}
%\vspace*{-15pt}
\label{fig:archi}
\end{figure}

IIoT systems adopt hierarchical architectures to simplify operation management and help with transparency in the workplace.
Usually, a hierarchical network consists of three tiers~\cite{bibli:myac}: backbone tier, distribution tier, and access tier, as shown in Fig.~\ref{fig:archi}.
The benefit of adopting the three-tier architecture is the clarity of roles performed by each tier.
This in turn separates the hardware features required and has already helped Cisco make networks more resilient and scalable\footnote{\url{https://www.ciscopress.com/articles/article.asp?p=474236}}.
However, lots of resilience concerns keep emerging as industrial systems become more complex.

%Resilience features should be well designed to span throughout all three tiers.
\subsubsection{Centralized Backbone Tier}
The backbone of IIoT systems generally works in a logically centralized manner.
A centralized control offers more opportunities for reducing costs, promoting security, and increasing transparency.
Its powerful computing and storage capabilities also enable optimal decision-making, which is based on the analysis of extensive historical information and globally collected real-time data.
The massive data, transferring towards the backbone, however, poses much strain to backhaul links, which needs meticulous employment of technologies such as software-defined network (SDN)~\cite{bibli:reroute,bibli:lyu} and network function virtualization (NFV)~\cite{bibli:NFV} to avoid potential congestion.
Besides, the centralized approach usually has strictly synchronized configuration requirements.
Any small configuration errors in one entity could be propagated immediately to the rest of the network, causing unbearable catastrophic failures.
Appropriate solutions like ring-shaped topology are needed for mitigating the influence of these failures.

\subsubsection{Decentralized Distribution Tier}
With the performance distributed across nodes (such as web servers, database servers, and load-balancers), a level of redundancy can be achieved at the distribution tier.
The decentralized manner enables backup routing scalability and is effective in avoiding complete network collapse.
Specifically, the fault of a node activates the rerouting process, which would only have a slight impact on neighboring nodes and overall performance~\cite{bibli:Globe,bibli:multipath}.
In large IIoT systems, operators can also use mobile sites to further enhance network robustness with low cost and high flexibility.
Interworking solutions among different nodes is essential for ensuring service continuity among nodes that belong to different parts of the system.
As multiple technologies and infrastructure from various suppliers will be put into use in IIoT, multi-vendor interoperability becomes more than ever critical to support the interconnection.

\subsubsection{Access Tier}
Access tier consists of outermost devices and sensors responsible for collecting and monitoring.
It is commonly accepted that resilience decreases when one moves from center to extremity~\cite{bibli:ENISA}.
On the outermost, components in the access tier such as base stations and sensors are generally of much less or even no resilience from the perspective of architecture design.
Any malfunction of a node at the access tier only affects itself and does less harm to other devices.
Therefore, in resilient architecture design attempts, more significance is attached to the upper two layers, especially the inherent interdependence between them.

\subsubsection{Interdependence Analysis}
The interconnectivity, technological dependency, and the data that fuels it between the backbone tier and distribution tier making IIoT a typical interdependent network~\cite{bibli:inter}.
As the number of devices and interconnections increases, IIoT is becoming more than ever fragile and is more likely to encounter cascading failure, especially in facing a malicious attack~\cite{bibli:multipath}.
For example, an attacker could maliciously manipulate a router to stop processing anything it henceforth receives and start flooding the system with a broadcast signal commanding other nodes to do the same, leading to an entire network breakdown.
Unfortunately, such a breakdown of the network is not easy to fix either.
If an IIoT does not have a traceability mechanism, then a late-induced failure cannot be fixed until the initial attacked node gets found by visiting all the interdependent parts in the system~\cite{bibli:trace}.
Given the interdependence in an IIoT system, it nearly equals checking all infrastructure in the network, which is infeasible considering IIoT's large scale.
To avoid such a disaster happening, it is critical for the IIoT system to exploit data redundancy to enhance information and identity verification.

\subsection{Data Redundancy}
In IIoT, numerous devices and components participate in the collection, transmission, and storage of data.
The completeness and accuracy of data remain a serious problem that needs to be addressed as the complexity of the system increases.
As an important solution, intentional data redundancy can be used to ensure information consistency and enhance system resilience.

\subsubsection{Data Collection}
Much of the data in IIoT is collected via sensors in the access tier that are error-prone and unreliable.
To prevent the system from losing critical data, multiple sensors can be used to collect the same data.
Therefore, the system could use alternative backup data for analysis when data from one source gets discarded or contaminated.
This type of data redundancy offers an extra layer of protection and reinforces the fault-tolerance of systems.
However, it may also increase the collected data size, putting more burden on subsequent data transmission and storage.
The number of data replications and system costs need to be well balanced depending on the essentiality of the data.

\subsubsection{Data Transmission}
Data redundancy can also be used to doubly check data and confirm its correctness after transmission, protecting data from cyberattacks and breaches~\cite{bibli:DT}.
On the one hand, algorithm-based fault tolerance approaches could be developed for matrix-based and signal processing applications such as matrix multiplication, matrix inversion, and lower-upper decomposition.
Given their software-level implementation, they are especially suitable for distributed networks where massive devices require credible information.
On the other hand, the same piece of data can be transmitted through separate paths to the destination node, which is especially helpful when interacting with managers, engineers, and others.
Although the cost burden of bandwidth and delay may increase by leveraging the data redundancy, unwanted activities like network collapse caused by interdependency could be reduced.

\subsubsection{Data Storage}
Data redundancy in storage (data cache) is a common occurrence in IIoT~\cite{bibli:storage}.
The duplicates of data are stored in different places for security and faster data access.
A distributed data storage environment could facilitate real-time monitoring and analysis but may augment the complexity of the database, making it more challenging to maintain.
A prompt and efficient caching algorithm is needed to decide the places for storing and the frequency for updating.

The redundancy provided by data is an important supplement to the topology-based routing redundancy.
Each network node has its own standalone configuration for data collection, transmission, and storage, relaxing the burden of topology-level fault tolerance and benefiting the rapid detection of faults in the system.

\subsection{Fault Detection}

Fault detection, including fault event detection, fault localization, and fault type determination, is critical to the quick repair of damaged systems~\cite{bibli:hardware}.
Basically, there are two types of faults: permanent and temporary faults.
The permanent fault, also called internal failure or primary failure, is caused by the breakdown of physical components, thus usually requires human intervention for recovery.
On the contrary, the temporary fault, without any hardware damage, could self-repair after a certain time interval.
Generally, the temporary fault is either caused by nodes' self-protection mechanisms like overheating protection or results from losing connection to the system giant component (GC).
As it is often a subsequent result of crucial connecting nodes' internal failure, the temporary fault caused by isolation from the GC is also known as secondary failure or external failure in cascading failure fields.

In IIoT, typically, two kinds of methods can be applied for fault detection: software and hardware methods.
Nodes with computational capability (e.g., edge computing nodes) can implement self-testing software to continuously monitor their own functional states (i.e., fluctuation of quality of service) and request assistance once a fault is found.
The consecutive state monitoring also provides an opportunity for profound techniques such as deep learning and data mining to be employed.
For software mechanisms, the speed of fault detection depends much on the signal processing procedure.
Conventional fault detection techniques using spectral analysis (e.g., Fourier) may fail to handle IIoT's non-stationary conditions, incipient faults, and noise interference environments.
To deal with this, specific signal processing designs need to be tailored.
For instance, a fast fault detection design could use a maximum likelihood estimator for fault characteristic computation, and then make automatic decisions by generalized likelihood ratio testing.
Besides that, principal component analysis and Kullback-Leibler divergence-based signal processing techniques are also advanced for early anomaly diagnosis~\cite{bibli:signal}.
However, the great deal of historic information storage and processing demands make the software mechanism inappropriate for some real-time services in IIoT.
For those time-critical services, the hardware mechanism~\cite{bibli:hardware}, with its faster-responding speed and more reliable function, would be a good alternative.
Nevertheless, dedicated hardware deployment is pretty expensive.
The trade-off between construction costs and system requirements needs to be well considered.

\subsection{Post-damage Response}
After faults are detected, rapid repair of the damaged network looms.
Repairing mechanisms may vary depending on the extent of the damage.
Local damage is the most common failure that often occurs in our daily life.
Sometimes, the damage is limited to only one single node or link.
In this case, the failure would not be propagated, only slightly influencing the performance of its nearby network.
To fix this type of damage, usually, service interruption and repair cost are the two most important things that need to be well balanced.
When it comes to multi-node failure scenarios, a mobile access point that served as a relay could be an economical way to ensure network connectivity.
Although the aforementioned methods could deal with most of the failures in IIoT, the ignorance of the global dynamic nature (e.g., data migration flows) and deployment restrictions (e.g., wired link layout) make them inappropriate for networks with large-scale damage.

IIoT is prone to large-scale damage, as it is an interdependent system comprising of different sub-networks like the power grid and edge computing sub-network.
For large-scale damage, the initially available repair resources (such as repairmen and replaceable equipment) are often limited.
How to maximize the usage of resources to satisfy the demand for performance recovery is an immediate problem.
Generally, we would believe that nodes with a high degree (hubs) and links with large betweenness centrality are worthy of priority restoration.
However, the intuition may not be true according to a study based on the data from Twitter~\cite{bibli:nature}.
Some "weak nodes" (nodes with lower degrees surrounded by hubs) may outrank the hubs in terms of recovery priority.
In IIoT, other than the topology, information of equipment details and migration flows in the network are also significant.
The consideration of those system details fosters the emergence of the network design problem~\cite{bibli:SR}, which is at least NP-complete and necessitates a reliable algorithm to obtain optimum within polynomial time.

\section{Promising Resilient Applications in IIoT}
The IIoT can be broadly categorized into three sectors~\cite{bibli:IIoT4}: extraction (e.g., intelligent agriculture), service (e.g., smart logistics), and manufacturing (e.g., smart factory).
Resilience acts as a pivotal part of almost every sector.
To illustrate that, we present some interesting areas in IIoT where enhancement methods can be applied to achieve the goals of network resilience.
Companioned with these great opportunities, challenges in each scenario are also discussed.

\subsection{Intelligent Agriculture}
Traditional agriculture is undergoing a considerable reform whose main driving force is IIoT technology.
Intelligent agriculture is the advanced stage of the farming process which integrates smart sensing, big data, and edge computing, aiming to tackle climate change, conserve water, and enhance productivity.
According to a report from Cision\footnote{\url{https://www.prnewswire.com/news-releases/global-smart-agriculture-market-to-reach-17-1-billion-by-2026--301323285.html}, by 2026, the total addressable market in global intelligent agriculture will reach 17.1 billion dollars.}
More connectivity and end-to-end services involved in intelligent agriculture will inevitably require the support of a resilient system.

In reality, resilience concerns occur in almost every domain of intelligent agriculture.
For instance, unmanned aerial vehicles (UAVs) used for crop monitoring need to be resilient to unwanted collecting and transmitting errors.
With the flexibility of mobility, UAVs can scan a vast area at low cost, and coordinate with multiple sensors to collect a wide range of information such as climate conditions, soil densification, and grain quantity.
The same pieces of data from the same district can be gathered by different UAVs and distributed stored in UAVs' data buffers to build redundancy.
Subsequently, multiple pieces of data are transmitted from UAVs to the edge computing server for analysis.
By leveraging the data redundancy, the impact of sampling deviation and transmission errors could be mitigated.
Therefore, farmers can have an accurate grasp of planting conditions and adopt optimal actions for plant growth.

\subsection{Smart Logistics}
Smart logistics refers to the integration of traffic management structuring and navigating traffic for optimal trip routes and vehicle usage by intelligent technologies.
Leveraging a large amount of collected data, it can organize, plan, control and execute the goods flow in an effective way.
Powered by the development of automated driving, smart logistics has been deemed the foundation of the manufacturing supply chain.
Although intelligent, vehicles in logistics networks may still encounter traffic congestion resulting from untimely schedules and malicious attacks.
Unfortunately, given the interdependency among vehicles and roadside units (RSUs), logistics networks are prone to cascading failures once a malfunction happens.
Therefore, strengthening the resilience of the logistics sector has emerged as a priority.

A resilient logistics network includes three-layer protection: vehicle, RSU, and network protection.
For vehicle protection, smart vehicles are equipped with onboard units, which could adaptively respond to traffic jams based on a pre-designed response set.
For example, a vehicle can reroute to a backup planning path and broadcast the congestion information to other vehicles by emitting a beacon signal.
Vehicles who receive the signal can further broadcast the congestion information through vehicle-to-vehicle (V2V) links to allow incoming counterparts to avoid problematic routes~\cite{bibli:cui}.
As the response set is predesigned by predictive events of historical data, there might be scenarios where no feasible response is available.
In this case, the RSU protection comes into play.
RSUs with more powerful computing ability and microscopic vision can assist the vehicle in rescheduling its route and forward the computational results to vehicles via vehicle-to-infrastructure (V2I) links.
In the worst cases, even RSUs are unable to find a feasible solution, the system then has to adopt network protection, in which a collaborative location platform serves as the backend component for traffic evacuation.
As a higher-level decision-maker, the platform is able to coordinate all the nodes in the system with a macroscopic view, thus eliminating the blind spots outside the reach of RSUs.
With global impact factors counted, optimal routing strategies can be made.
However, such optimal decisions are usually time-consuming and can only be applied to large-scale congestion mitigation.

\subsection{Smart Factory}
\begin{figure}
\centering{\includegraphics[width=1\linewidth]{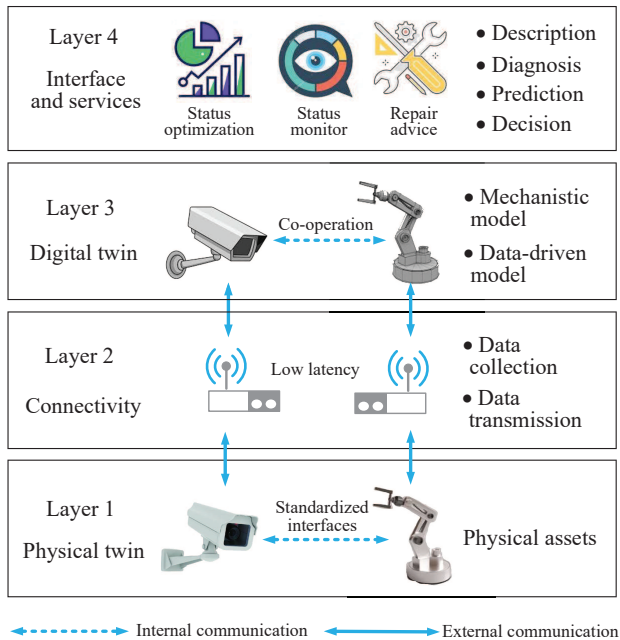}}
\caption{Role of communication in achieving a resilient production line with digital twins.}
%\vspace*{-15pt}
\label{fig:DT}
\end{figure}

As more transformative technologies develop, manufacturing plants, which are the main scene of Industry 4.0, are brought into the era of smart factories.
New features such as digital twins, automated mobile robots (AMRs), big data, and artificial intelligence continue to push the limits of production efficiency, transferring the production environment to a fully digital stage.
However, being fully digital is also risky: the downtime of asset components could be extremely expensive since inherent interdependence would spread a tiny failure to the entire network.
Resilience, as the supporting block of smart factories, must be put on the manufacturing agenda.

Different from the scenarios in smart logistics and automated driving, AMRs in smart factories could use more context information to enhance their resilience~\cite{bibli:EMM}, thanks to the regular routes within a warehouse or facility.
Context information such as mobility speed and obstacle location could be transmitted to and stored at edge computing nodes for analysis, or even be forwarded to the AI-powered cloud for advanced learning behaviors~\cite{bibli:Learn}.
With multiple input data from AMRs, collecting and transmitting errors could be corrected to form an accurate context-aware mobility environment.
AMR itself, with relatively powerful computing ability, could also employ self-testing software to monitor its working status.
Therefore, avoid breakdowns or power outages during the execution of tasks.

Beyond automation and computing, the smart factory also creates a digital twin as a virtual model to reflect resilience dynamics.
For instance, in a production line, as illustrated in Fig.~\ref{fig:DT}, ubiquitous sensors embedded in infrastructure collect a vast amount of diverse real-time data from the physical twin and environment.
These data are then transmitted to the cloud or edge nodes through a low-latency connection in the shape of 5G to form the virtual digit twin.
Any small malfunctions within the production line could be instantaneously reflected on its digital counterpart, and rapidly informed to engineers or managers for handling.
Engineers can then take quick responses to head off the failure from escalating.
Besides, a digital twin can stimulate the risks and vulnerabilities hiding in the system.
When a segment failure or non-optimal performance is predicted, manufacturers can get ahead of problems and optimize before the failure causes side effects to the bottom line.
To achieve the true potential of digit twin, the communication interoperability between components would require standardized interfaces and co-operation ability, which is pretty hard due to the heterogeneity (e.g., diverse devices from different companies with different purposes) and the fast-evolving nature of IIoT.

\vspace*{8pt}
\section{Conclusions}
The maturity level of the current resilient frameworks is in marked contrast to the dependence on and complexity of IIoT networks as a whole.
A widely accepted set of well-defined resilient terms from a communication perspective is essential yet still lacking.
In this paper, we have presented an attempt to provide a systematic overview of resilience-related concepts and metrics in IIoT systems.
It puts together work that has been done in the areas of redundancy, robustness, recoverability, and specific taxonomy research under the single umbrella of resilient communication.
We have also discussed the standardization efforts for the realization of resilient IIoT.
Besides, key resilience concerns that stem from the design, deployment, and operation in industrial communications are summarized and discussed.
Furthermore, promising resilience applications in industrial sectors are presented, which highlight the wide range of values of resilient research and motivate more in-depth future works on the subject.

% Can use something like this to put references on a page
% by themselves when using endfloat and the captionsoff option.
\ifCLASSOPTIONcaptionsoff
  \newpage
\fi

% trigger a \newpage just before the given reference
% number - used to balance the columns on the last page
% adjust value as needed - may need to be readjusted if
% the document is modified later
%\IEEEtriggeratref{8}
% The "triggered" command can be changed if desired:
%\IEEEtriggercmd{\enlargethispage{-5in}}

% references section

% can use a bibliography generated by BibTeX as a .bbl file
% BibTeX documentation can be easily obtained at:
% http://mirror.ctan.org/biblio/bibtex/contrib/doc/
% The IEEEtran BibTeX style support page is at:
% http://www.michaelshell.org/tex/ieeetran/bibtex/
%\bibliographystyle{IEEEtran}
% argument is your BibTeX string definitions and bibliography database(s)
%\bibliography{IEEEabrv,../bib/paper}
%
% <OR> manually copy in the resultant .bbl file
% set second argument of \begin to the number of references
% (used to reserve space for the reference number labels box)

\bibliographystyle{IEEEtran}
%% argument is your BibTeX string definitions and bibliography database(s)
\bibliography{Ref-MEC-TWC}

\end{document}